\documentclass[aps,twocolumn,groupedaddress,amsmath,amssymb,nofootinbib]{revtex4-1}
\usepackage{ulem}
\usepackage{amsmath, dsfont}
\usepackage{amssymb}
\usepackage[dvips]{graphics}
\usepackage{epsfig}
\usepackage{calc}
\usepackage{amsfonts}
\usepackage{graphicx}
\usepackage[ansinew]{inputenc}
\usepackage{xcolor}

\def\vp{\varphi}
\def\al{\alpha}
\def\b{\beta}
\DeclareMathOperator{\arcosh}{arcosh}
\DeclareMathOperator{\arsinh}{arsinh}

\begin{document}
\setcounter{page}{1}

\title[]{{Planar black holes configurations and shear viscosity in arbitrary dimensions with shift and reflection symmetric
scalar-tensor theories }}

\author{Mois\'es Bravo-Gaete}\email{mbravo-at-ucm.cl}
\affiliation{Facultad de Ciencias B\'asicas, Universidad Cat\'olica del Maule, Casilla 617,
Talca, Chile.}

\author{M. M. Stetsko}\email{mstetsko-at-gmail.com}
\affiliation{Department for Theoretical Physics, Ivan Franko National University of Lviv, 12 Drahomanov Str., Lviv, UA-79005, Ukraine}

\begin{abstract}
In higher dimensions, we explore planar hairy black hole configurations for a special subclass of the
Horndeski theory, \textcolor{black}{defined by two coupling functions depending on the kinetic term and enjoying shift symmetry and reflection symmetry.} For this analysis, we derive a set of new solutions, given by \textcolor{black}{time-dependent as well as time independent scalar field configurations.} Additionally, we calculate their thermodynamic \textcolor{black}{quantities} by using Wald formalism, \textcolor{black}{satisfying} the First Law of Thermodynamics
as well as a Smarr relation. Together with the above, the Wald procedure allows \textcolor{black}{us} to compute the
shear viscosity, showing that for a suitable choice of the coupling functions the Kovtun-Son-Starinets
bound \textcolor{black}{is violated}.

\end{abstract}

\maketitle
%\tableofcontents
\newpage

%%%%%%%%%%%%%%%%%%%%%%%%%%%%%%%%%%%%%%%%%%%%%%%%%%%%%%%%
\section{Introduction}
\label{intro}
%%%%%%%%%%%%%%%%%%%%%%%%%%%%%%%%%%%%%%%%%%%%%%%%%%%%%%%%
{General Relativity (GR) is without a doubt a very successful standard model of gravity. Nevertheless, astrophysical discoveries such as the accelerated expansion of the Universe \cite{Riess:1998cb,Perlmutter:1998np} and the recent first detection of gravitational waves \cite{Abbott:2016blz}, have yielded the motivation to study theories of gravity beyond GR.}

There are many ways to construct modifications of the GR, one of them is \textcolor{black}{to introduce}  new degrees of freedom, given by scalar fields, and denominated as {scalar-tensor-theories}. In particular, G. Horndeski in the seventies formulated a four-dimensional theory defined by the metric $g_{\mu \nu}$ and a scalar field $\phi$, as well as their derivatives, where the equations of motion are at most of the second-order \cite{Horndeski:1974wa}. \textcolor{black}{This peculiarity makes the Horndeski theory to be a {healthy one}, because it does not have ghosts or instabilities caused by higher orders derivatives in the equations of motion.}

{An intuitive way to construct the Horndeski theory action follows from the Galileon theory \cite{Deffayet:2009mn,Charmousis:2014mia}, with the introduction of four functions dependent of the metric $g_{\mu \nu}$, the scalar field $\phi$ and the kinetic term $X:=-\frac{1}{2} g^{\mu \nu} \partial_{\mu} \phi \partial_{\nu} \phi$ which reads \cite{Kobayashi:2011nu}
\begin{eqnarray}\label{actionhorn}
&&S[g_{\mu \nu},\phi] = \int d^{4} x \sqrt{-g}
\Big[G_2\left(X,\phi\right)-G_3\left(X,\phi
\right)\Box\phi\nonumber\\
&&+G_{4 X}\,\Big(\left(\Box\phi\right)^{2}
-\left(\nabla_{\mu}\nabla_{\nu}\phi\right)^{2}\Big)
+R\, G_{4}(X,\phi) \nonumber\\
&&-\frac{1}{6}\,G_{5X}
\Big(\left(\Box\phi\right)^{3}-3\,\Box\phi\left(\nabla_\mu \nabla_\nu \phi\right)^{2}
+2\left(\nabla_\mu \nabla_\nu \phi\right)^{3}\Big)\nonumber\\
&&+G_{\mu\nu}\nabla^{\mu}\nabla^{\nu}\phi\,
 \textcolor{black}{G_{(5)}}\left(X, \phi\right)\Big],
\end{eqnarray}
where we define
\begin{eqnarray*}
G_{n X}&:=&\frac{\partial G_{n}}{\partial X},\qquad
n=\{2,3,4,5\},\\
\left(\nabla_\mu \nabla_\nu \phi\right)^{2} &:=& \left(\nabla_{\mu}\nabla_{\nu}\phi\right)\left(\nabla^{\mu}\nabla^{\nu}\phi\right),\\
\left(\nabla_\mu \nabla_\nu \phi\right)^{3} &:=&
\left(\nabla_{\mu}\nabla_{\nu}\phi\right)
\left(\nabla^{\mu}\nabla^{\rho}\phi\right)\left(\nabla_{\rho}\nabla^{\nu}\phi\right),
\end{eqnarray*}
with $R$ and $G_{\mu \nu}$ being the scalar curvature and the Einstein tensor respectively.}

{On the other hand, one of the peculiarities of planar black holes is \textcolor{black}{their relation to an ideal fluid given by the} gravity/gauge duality \cite{Maldacena:1997re,Gubser:1998bc,Witten:1998qj}. Within this scenario, it is possible to compute the well-known ratio between the shear viscosity $\eta$ and the entropy density ${s}$, allowing us to check the conjecture about a universal bound, known as  the Kovtun-Son-Starinets (KSS) bound, which reads \cite{Policastro:2001yc,Son:2002sd,Kovtun:2003wp,Kovtun:2004de} 
\begin{equation}\label{KSS}
\frac{\eta}{s}\geq \frac{1}{4 \pi}, 
\end{equation}
being demonstrated in a variety of gravity theories (see for example \cite{Buchel:2003tz,Buchel:2004qq,Benincasa:2006fu,Landsteiner:2007bd}),
where the shear viscosity can be obtained by effective coupling constants of the transverse graviton on the location of the event horizon, via the membrane paradigm \cite{Iqbal:2008by}, and corroborated by the Kubo formula \cite{Cai:2008ph,Cai:2009zv}. Recently, constructing a Noether charge with a suitable choice of a space-like Killing vector, \textcolor{black}{ and following the Wald formalism \cite{Wald:1993nt,Iyer:1994ys} }, the  \textcolor{black}{$\eta/s$ ratio} was calculated by using the infrared data on the black hole event horizon \cite{Fan:2018qnt}, greatly simplifying the steps in comparison with the previous procedure.}

Nevertheless, in recent years it has been shown with specific examples that the bound can be violated. In fact, we can mention \textcolor{black}{gravity} theories such as the Einstein-Hilbert Gauss-Bonnet model \cite{Kats:2007mq,Brigante:2007nu}  as well as a particular truncation of the Horndeski theory (\ref{actionhorn}) \textcolor{black}{\cite{Feng:2015oea,Brito:2019ose}.}

{In particular, in the present work, we are interested in the study of a subclass of the action (\ref{actionhorn}) based on the work developed in \cite{Kobayashi:2014eva}, where the theory in $D$-dimensions takes the form
\begin{eqnarray}\label{action}
S[g_{\mu \nu},\phi]=\int d^{D}x \sqrt{-g} \mathcal{L},\label{action}
\end{eqnarray}
and the Lagrangian is expressed as
\begin{eqnarray}
{\cal L}=G_2 +G_4 R+G_{4X}\left[\left(\Box\phi\right)^2
-\left(\nabla_\mu\nabla_\nu\phi\right)^2\right], \label{Lagrangian}
\end{eqnarray}
where now $G_2$ and $G_4$ are arbitrary \textcolor{black}{functions} of the kinetic term $X$ and, as before, $G_{4X}:=\partial
G_4/\partial X$. The corresponding equations of \textcolor{black}{motion are of} the following form:
\begin{eqnarray}
{\cal E}_{\mu\nu}:=
\frac{2}{\sqrt{-g}}\frac{\delta\left(\sqrt{-g}{\cal
L}\right)}{\delta g^{\mu\nu}} =0,\label{emunu}
\end{eqnarray}
\begin{eqnarray}
\textcolor{black}{{\cal E}_{\phi}:=}\nabla_\mu J^\mu=0,\label{seom}
\end{eqnarray}
where
\begin{eqnarray}
J^\mu&:=&-G_{2X}\nabla^\mu\phi +2G_{4X}G^{\mu\nu}\nabla_\nu\phi
\nonumber\\&& -G_{4XX}\left[\left(\Box\phi\right)^2
-\left(\nabla_\mu\nabla_\nu\phi\right)^2\right]\nabla^\mu\phi
\nonumber\\&& -2G_{4XX}\left(\Box\phi\nabla^\mu
X-\nabla^\mu\nabla^\nu\phi\nabla_\nu X\right), \label{eq:jmu}
\end{eqnarray}
while the equations with respect to the metric
${\cal E}_{\mu\nu}$ are reported in the Appendix. Within this theory, we will focus on black holes in arbitrary dimensions with planar base manifold for the event horizon, thermodynamics of these configurations will be also examined. In addition, we compute the shear viscosity $\eta$, and to perform this task in the present paper we will utilize the formalism developed in \cite{Fan:2018qnt}.}

The rest of the paper is organized as follows. In \textcolor{black}{the} Section \ref{time-dependent} we explore planar black holes with a linear time-dependent scalar field $\phi$, giving a general solution for some particular cases for the functions $G_2$ and $G_4$. In the Section \ref{time-independent} the time-independent case is analyzed, whereas its thermodynamic is studied in \textcolor{black}{the} Section \ref{sec.termo}.  In \textcolor{black}{the} Section \ref{viscosity}, the shear viscosity is computed, where the \textcolor{black}{$\eta/s$ ratio} is obtained, and a condition on the functions $G_2$ and $G_4$ is found where the KSS bound can be violated. Finally, \textcolor{black}{the} Section \ref{conclusions} is devoted to our conclusions and discussions.

%%%%%%%%%%%%%%%%%%%%%%%%%%%%%%%%%%%%%%%%%%%%%%%%%%%%%%%%%%%
{\section{Derivation of the solution with a linear time-dependent scalar field}\label{time-dependent}}
%%%%%%%%%%%%%%%%%%%%%%%%%%%%%%%%%%%%%%%%%%%%%%%%%%%%%%%%%%%

One of the peculiarities of scalar fields in scalar-tensor theories of Horndeski type (\ref{action})-(\ref{Lagrangian}) is the existence of time-dependent configurations compatible with the gravitational sector. In particular, \textcolor{black}{in the following studies, the
ansatz for the metric will be}
\begin{eqnarray}
ds^2=-h(r)\,dt^2+\frac{dr^2}{f(r)}+r^2 \sum_{i=1}^{D-2} d x_{i}^2,
\label{metricd}
\end{eqnarray}
\textcolor{black}{and following the steps performed in \cite{Babichev:2013cya, Kobayashi:2014eva},  the scalar field reads:}
\begin{equation}\label{phi-time}
\phi(r,t)=\vp(r)+qt,
\end{equation}
where $q$ is a constant. It is worth pointing out that this structure for the scalar field has been successfully applied with other kinds of topologies for the event horizon (see for example \cite{Kobayashi:2014eva, Babichev:2013cya,Anabalon:2013oea,Rinaldi:2012vy,Bravo-Gaete:2014haa}). Since \textcolor{black} {that} the scalar field $\phi$ \textcolor{black}{is time-dependent}, the kinetic term $X$ takes the form:
\begin{equation}\label{X}
X=\frac{1}{2}\left(\frac{q^2}{h}-f(\vp')^2\right).
\end{equation}
Imposing a condition on the radial part of the current $J^{r}=0$, we obtain the following relation for the metric functions $f$ and $h$:
\begin{widetext}
\begin{equation}
f=-\frac{G_{2X}r^{D-2}h}{(D-2)\left[ {\cal G}_{X} (r^{D-3}h)'+(D-3)q^2r^{D-4}G_{4XX}\right]},\label{fh_time}
\end{equation} 
\end{widetext}
{where $(')$ denotes the derivative with respect \textcolor{black}{to} the radial coordinate $r$ and 
\begin{eqnarray}\label{functionG}
\mathcal{G}(X)=\,\big(G_{4}-2\,X\,G_{4X}\big), \qquad \mathcal{G}_{X}:=\partial \mathcal{G}/\partial X.
\end{eqnarray}
It should be pointed out that for \textcolor{black}{the time-dependent scalar field (\ref{phi-time})}, in addition to ``diagonal'' gravitational field equations (\ref{emunu}) there is an ``off-diagonal'' term which corresponds to $(t,r)-$component, where the corresponding field equation takes the form:
\begin{eqnarray}
&&{\cal E}_{tr}:=\left[-\frac{G_{2X}}{2}+G_{4X}\frac{(D-2)f}{2r^{D-2}h}(r^{D-3}h)'+G_{4XX}\right. \nonumber \\
&&\left.\times\frac{(D-2)f}{2rh}\left(\frac{h'}{h}q^2+\frac{(r^{D-3}h)'}{r^{D-3}}f(\vp')^2\right)\right]q\vp'=0.  \label{tr_grav}
\end{eqnarray}
\textcolor{black}{This latter equation, together with (\ref{X}),} also gives rise to the relation (\ref{fh_time}), and it is worth noting that this fact reflects the consistency of our procedure.
\textcolor{black}{Taking into account the relation (\ref{fh_time}), we can recast the  $(r,r)-$component of the equations of motion (\ref{emunu}) in the following form}
\begin{eqnarray}
(r^{D-3}h)'\partial_{X}(G_2{\cal G})+&&\nonumber\\
(D-3)q^2r^{D-4}\partial_{X}(G_2G_{4X})=0.\label{eq_Grr}&&
\end{eqnarray}
Having combined the equation (\ref{eq_Grr}) and the relation for  metric functions (\ref{fh_time}), we can also derive the following relation:
\begin{equation}
f=-\frac{r^2}{(D-2)(D-3)q^2}\frac{\partial_{X}(G_2{\cal G})}{\partial_{X}(G_4G_{4X})}h,\label{fh_new}
\end{equation}
where we note that it \textcolor{black}{cannot be applied} to a time-independent scalar field (\textcolor{black}{this is} $q=0$).

\textcolor{black}{Using the {equations  (\ref{fh_time}),  (\ref{eq_Grr}) and (\ref{fh_new})},} the $(t,t)-$component of the Einstein equations (\ref{emunu}) acquires the structure:
\begin{eqnarray}
-\frac{(D-2)h}{2r{\cal G}} \left(\frac{f}{h}{\cal G}^2\right)'=0,\label{G_tt}
\end{eqnarray}
and it follows immediately that:
\begin{equation}
\frac{f}{h}{\cal G}^2=C,\label{int_G_tt}
\end{equation}
where $C$ is an integration constant. \textcolor{black}{Here it is important to note that we do not impose any asymptotic behavior at the beginning, but it follows from the solutions we will obtain below. Finally, combining the solution (\ref{int_G_tt}) together with the relation (\ref{fh_new}), we obtain}
\begin{equation}
-\frac{r^2}{(D-2)(D-3)q^2}\frac{\partial_{X}(G_2{\cal G})}{\partial_{X}(G_4G_{4X})}=\frac{C}{{\cal G}^2}.\label{X_rel}
\end{equation}\\
This relation demonstrates clearly that kinetic term $X=X(r)$ can be derived algebraically for given functions $G_2$ and $G_4$, \textcolor{black}{and it is in complete agreement with the result obtained previously for the four-dimensional case \cite{Kobayashi:2014eva}. Having obtained the explicit form of the function $X(r)$, one can use the equation (\ref{eq_Grr}) to find the metric function $h(r)$, and finally using the relation (\ref{fh_new}) we derive the metric function $f(r)$.} We also point out that the constant $C$ in the relation (\ref{int_G_tt}) should be positive ($C>0$), because the metric functions $h(r)$ and $f(r)$ should be of the same sign in any point on their domains of variation. In the following lines, we set $C=1$ without any loss of generality of conclusions.

\textcolor{black}{Since any specific condition has not been imposed on the functions $G_2$ and $G_4$, we can choose them freely and try to find explicit structures for the metric functions $h(r)$ and $f(r)$, as well as the explicit expression for the kinetic term $X(r)$.} Here below we examine \textcolor{black}{a} few cases for the functions $G_2$ and $G_4$ for which it is possible to obtain explicit expressions of the metric functions, at least in a relatively simple \textcolor{black}{form. Before we start considering some particular cases, we would like to stress that the only equation which has not been used explicitly when we derived the relations (\ref{fh_new}), (\ref{int_G_tt}) and (\ref{X_rel}), is the $(x_i,x_i)-$component from the field equations (\ref{emunu}).}  Therefore, we might expect that this equation gives rise to some conditions which should be imposed \textcolor{black}{on} the functions $G_2$ and $G_4$, being written as follows:
\begin{widetext}
\begin{equation}
{\cal G}\left[\frac{1}{2}\sqrt{\frac{f}{h}}\left(\sqrt{\frac{f}{h}}h'\right)'+\frac{(D-3)}{2r^{D-3}h}(r^{D-4}fh)'\right]-\frac{G_2}{2}+{\cal G}'\left(\frac{h'}{2h}+\frac{D-3}{r}\right)f+\frac{(D-3)q^2}{2r^{D-3}G_{4X}}\left(\frac{r^{D-4}fG^2_{4X}}{h}\right)'=0.
\end{equation}
\end{widetext}
If we now use the relation (\ref{int_G_tt}), the latter equation \textcolor{black}{can} be simplified to
\begin{equation}\label{G_22}
\frac{(r^{D-3}h)''}{r^{D-3}{\cal G}}-G_2+\frac{(D-3)q^2}{r^{D-3}G_{4X}}\left(\frac{r^{D-4}G^2_{4X}}{{\cal G}^2}\right)'=0.
\end{equation}
It is worth noting that \textcolor{black}{written above equation can be applied} to the time-independent field, namely in this case $q=0$, and the equation can be cast \textcolor{black}{as follows:}
\begin{equation}\label{G_22-1}
(r^{D-3}h)''-r^{D-3}{\cal G}G_2=0.
\end{equation}
\textcolor{black}{In the following, we will use the equation (\ref{G_22}) to check whether some additional conditions for $G_2$ and $G_4$ appear, allowing us to obtain explicit expressions for $h(r)$ and $f(r)$.}\\

%%%%%%%%%%%%%%%%%%%%%%%%%%%%%%%%%%%%%
\subsection*{Case A}
%%%%%%%%%%%%%%%%%%%%%%%%%%%%%%%%%%%%%

\textcolor{black}{As a first case, we assume that both $G_2$ and $G_4$ are linear functions of $X$:}
\begin{equation}\label{case_1}
G_2=\al+\b X,\quad G_4=\xi+\gamma X,
\end{equation}
where $\al$, $\b$, $\xi$ and $\gamma$ are constants. \textcolor{black}{By using the relation (\ref{X_rel})  together with $\beta=-\gamma \alpha / \xi$, we obtain:}
\begin{equation}\label{X_c1}
X=\frac{1}{\gamma}\left[\xi-\left(\frac{\gamma \xi (D-2)(D-3)}{2\al}\right)^{1/3}\left(\frac{q}{r}\right)^{2/3}\right].
\end{equation}
\textcolor{black}{Now, taking into account the equation (\ref{eq_Grr}), one can obtain the explicit structure for the metric function $h(r)$, which takes the following form:}
 \begin{equation}\label{h_c1}
h(r)=-\frac{M}{r^{D-3}}-\left(\frac{2(D-3)^2\al\gamma^2q^4}{\xi (D-2)}\right)^{1/3}\frac{3\,r^{2/3}}{(3D-7)},
\end{equation}
\textcolor{black}{where $M$ is a positive integration constant.} In order to have a black hole solution the metric function $h(r)$ should be positive at least for sufficiently large $r$ \textcolor{black}{, consequently,} it means that the second term in the relation (\ref{h_c1}) has to be positive and it takes place if $\al/\gamma<0$ with $\xi>0$.
To derive \textcolor{black}{the metric function $f(r)$, } according to the relation (\ref{int_G_tt}), one also needs \textcolor{black}{the} explicit relation for the function ${\cal G}$ \textcolor{black}{, being obtained via the relations (\ref{functionG}) and (\ref{X_c1}), given by} 
\begin{equation}\label{gsq_c1}
{\cal G}^2=\left(\frac{\gamma \xi (D-2)(D-3)q^2}{2\al}\right)^{2/3}\frac{1}{r^{4/3}}.
\end{equation}
Substituting the functions (\ref{h_c1}) and (\ref{gsq_c1}) into the equation (\ref{G_22}), we can check that this equation is satisfied as an identity without any conditions on the parameters of the functions $G_2$ and $G_4$. Here we would also like to emphasize that a similar conclusion can be made regarding all of the \textcolor{black}{chosen forms of $G_2$, $G_4$ and $h(r)$ that we will consider below}. Just for completeness, after a redefinition of the coupling constants $\gamma,\alpha$, and $\xi$, as well as \textcolor{black}{for the integration constant $M$, we recover the asymptotically Lifshitz black hole in arbitrary dimensions found in \cite{Bravo-Gaete:2013dca}.} 

%%%%%%%%%%%%%%%%%%%%%%%%%%%%%%%%%%%%%
\subsection*{Case B}
%%%%%%%%%%%%%%%%%%%%%%%%%%%%%%%%%%%%%

Now we suppose that \textcolor{black}{the functions $G_2$ and $G_4$ are}
\begin{equation}
G_2=\al(\xi-\gamma X)^k, \quad G_4=\xi+\gamma X,
\end{equation}
\textcolor{black}{and using the equation (\ref{X_rel})} we obtain
\begin{equation}\label{X_c2}
X=\frac{1}{\gamma}\left[\xi-\left(\frac{(D-2)(D-3)\gamma q^2}{(k+1)\al}\right)^{\frac{1}{k+2}}r^{-2/(k+2)}\right],
\end{equation}
\textcolor{black}{where $k\neq -1$ and $k\neq -2$, and the} equation (\ref{eq_Grr}) gives rise to the following form of the metric function $h(r)$:
\begin{equation}\label{h_c2}
h(r)=-\frac{M}{r^{D-3}}-\frac{k(k+2)(D-3)\gamma q^2}{(k+1)[(k+2)(D-3)+2]}\frac{r^{\frac{2}{k+2}}}{\bar{A}},
\end{equation}
where
\begin{eqnarray*} 
\bar{A}=\left[\frac{(D-2)(D-3)\gamma q^2}{\al(k+1)}\right]^{1/(k+2)},
\end{eqnarray*}
with $M$ a \textcolor{black}{positive} integration constant.}
Finally, \textcolor{black}{via the equation (\ref{G_22-1}) we obtain}
{\begin{equation}\label{gsq_c2}
{\cal G}^2={\bar{A}}^2r^{-4/(k+2)},
\end{equation}
\textcolor{black}{allowing us to obtain the expression for the metric function $f(r)$ by using the relation (\ref{int_G_tt}).}

One can easily check that if $k=1$, and after a redefinition of the constants, the relations (\ref{X_c2})-(\ref{gsq_c2})} are reduced to the corresponding relations obtained for the previously examined \textcolor{black}{situation ({Case A}), corresponding to asymptotically Lifshitz black holes in higher dimensions.}

%%%%%%%%%%%%%%%%%%%%%%%%%%%%%%%%%%%%%
\subsection*{Case C}
%%%%%%%%%%%%%%%%%%%%%%%%%%%%%%%%%%%%%

\textcolor{black} {For} this situation, the functions $G_2$ and $G_4$ are chosen as
\begin{equation}\label{case_C}
G_2=\al+\b X^k,\quad G_4=\xi+\gamma\sqrt{X},
\end{equation}
\textcolor{black}{and} the relation (\ref{X_rel}) allows us to arrive at
\begin{equation}
X=\left[\frac{(D-2)(D-3)\gamma q^2}{4k\b\xi^2}\right]^{\frac{2}{2k+1}}r^{-\frac{4}{2k+1}}
\end{equation}
\textcolor{black}{where} $k\neq-\frac{1}{2}$. Using the equation (\ref{eq_Grr}) we obtain the metric function $h(r)$, which reads:
\begin{eqnarray}\label{h_c3}
\nonumber &&h(r)=-\frac{M}{r^{D-3}}+\frac{(D-3)\gamma q^2}{2k\xi\b}\left[\frac{\al}{2(D-1)}\tilde{A}^{-\frac{2k+1}{2}}r^2\right.\\
&&\left.-\frac{(4k^2-1)\b}{2\big[(D-3)(2k+1)+2\big]}\tilde{A}^{-\frac{1}{2}}r^{\frac{2k}{2k+1}}\right],
\end{eqnarray}
where, \textcolor{black}{as before, $M$ is a positive integration constant and} we denoted for simplicity:
\begin{eqnarray*}
\tilde{A}=\left[\frac{\gamma}{4k\b\xi^2}(D-2)(D-3)q^2\right]^{2/(2k+1)}.
\end{eqnarray*}
It is easy to check that:
\begin{equation}
{\cal G}^2=\xi^2,
\end{equation}
and as a consequence if we impose {$\xi=1$}, we have that $f(r)=h(r)$. \textcolor{black}{Here is important to note that for this case we have and (A)dS configuration depending on the sign of 
$$\frac{\al (D-3)\gamma q^2}{2(D-1) 2k\xi\b}\tilde{A}^{-\frac{2k+1}{2}},$$
together with $2k/(2k+1)<0$. For the sake of completeness}, starting with $k=1/2$ from (\ref{case_C}) we note that the third term in the relation (\ref{h_c3}) disappears and the metric functions take the very simple form:
\begin{equation}\label{ds_c3}
f(r)=h(r)=-\frac{M}{r^{D-3}}+\frac{\al}{(D-1)(D-2)}r^2.
\end{equation}
Therefore, for this particular case, we completely recover the \textcolor{black}{(A)dS} situations of solutions known from the standard General Relativity gravity. We also point out that the solution (\ref{ds_c3}) represents a black hole if we impose $\al>0$.

%%%%%%%%%%%%%%%%%%%%%%%%%%%%%%%%%%%%%
\subsection*{Case D}
%%%%%%%%%%%%%%%%%%%%%%%%%%%%%%%%%%%%%

Now, for an election of the functions given by
\begin{equation}\label{case_D}
G_2=\al X^2\left(1+\gamma X^2\right)^{3/4},\quad G_4=\xi\left(1+\gamma X^2\right)^{1/4},
\end{equation}
the equation (\ref{X_rel}) gives rise to:
\begin{equation}\label{X_c4}
X=-\frac{\gamma}{4\al\xi}\frac{(D-2)(D-3)q^2}{r^2},
\end{equation}
while that the metric function $h(r)$ takes the form:
\begin{equation}\label{h_c4}
h(r)=-\frac{M}{r^{D-3}}+\frac{3\gamma^2(D-2)(D-3)^2}{16\al\xi(D-5)}\frac{q^4}{r^2},
\end{equation}
\textcolor{black}{where from the equation (\ref{G_22-1}) we have
\begin{equation}
{\cal G}^2=\xi^2\left[1+\frac{\gamma^3(D-2)^2(D-3)^2}{16\al^2\xi^2}\frac{q^4}{r^4}\right]^{-\frac{3}{2}},
\end{equation}
while that the metric function $f(r)$ is obtained via the equation  (\ref{int_G_tt}). Here it is important to note that, unlike the previous cases, this solution enjoys other asymptotic behavior, where to obtain a black hole, the presence of the constant $q$ as well as a positive integration constant $M$ are providential.} It is worth noting that if $D=5$ the second term in (\ref{h_c4}) instead of inverse quadratic dependence will have $\sim \ln(r)/r^2$ character, allowing us to study the number of the locations of the event horizon by using  Lambert W functions \cite{BravoGaete:2019rci}. \textcolor{black}{It should be also pointed out that to have a black hole we have to impose $D\geqslant 5$, because if $D=4$ the terms in (\ref{h_c4}) reverse their roles, and for the particular case $D=3$ the function (\ref{h_c4}) keeps the first term only.}

%%%%%%%%%%%%%%%%%%%%%%%%%%%%%%%%%%%%%
\subsection*{Case E}
%%%%%%%%%%%%%%%%%%%%%%%%%%%%%%%%%%%%%

\textcolor{black}{For this case,} let us take the functions $G_2$ and $G_4$ in the following form:
\begin{equation}\label{case_E}
G_2=\al\left(1+\gamma X^2\right)^{5/4}, \quad G_4=\xi\left(1+\gamma X^2\right)^{1/4},
\end{equation}
\textcolor{black}{where} from the equation (\ref{X_rel}) we obtain
\begin{equation}\label{X_c5}
X=-\left[\frac{4\al^2\xi^2}{(D-2)^2(D-3)^2 q^4}r^4-\gamma\right]^{-1/2},
\end{equation}
and it follows immediately that the condition on the coordinate $r$ that should be imposed in order to provide the function $X(r)$ to be real is {$r^4\geqslant \gamma(D-2)^2(D-3)^2q^4/(4\al^2\xi^2)$}, but we assume that these parameters are chosen in such a way that this condition is fulfilled in the outside of the event horizon of the black hole that we are going to find here. Using (\ref{X_c5}), we are able to find that the metric function $h(r)$ is given by
\begin{eqnarray}\label{h_c5}
\nonumber h(r)=-\frac{M}{r^{D-3}}+\frac{(D-3)q^2}{2r^{D-3}}\times\\\nonumber\int dr r^{D-4}\left[\left(\frac{4\al^2\xi^2}{(D-2)^2(D-3)^2q^4}r^4-\gamma\right)^{\frac{1}{2}}\right.\\\left.+2\gamma\left(\frac{4\al^2\xi^2}{(D-2)^2(D-3)^2q^4}r^4-\gamma\right)^{-\frac{1}{2}}\right].
\end{eqnarray}
In general, the integrals in the written above relations cannot be written in terms of elementary functions. But {it can} be represented in \textcolor{black}{terms of} hypergeometric functions as:
\begin{eqnarray}
\nonumber h(r)=-\frac{M}{r^{D-3}}+\frac{\al\xi}{(D-1)(D-2)}r^2\times\\\nonumber{_{2}}F_{1}\left(-\frac{1}{2},\frac{1-D}{4};\frac{5-D}{4};\frac{\gamma(D-2)^2(D-3)^2q^4}{4\al^2\xi^2 r^4}\right)+\\\nonumber\frac{\gamma(D-2)(D-3)^2q^4}{2\al\xi(D-5)r^2}\times\\{_{2}}F_{1}\left(\frac{1}{2},\frac{5-D}{4};\frac{9-D}{4};\frac{\gamma(D-2)^2(D-3)^2q^4}{4\al^2\xi^2 r^4}\right).\label{metr_hypergeo}
\end{eqnarray}
The structure of the written above solution might be a bit complicated, but for instance, for large $r$ ($r\rightarrow +\infty$) the asymptotic form of this function is as follows:
\begin{eqnarray}\label{asymp_h_c5}
h(r)&\simeq& -\frac{M}{r^{D-3}}+\frac{\al\xi}{(D-1)(D-2)}r^2\nonumber\\
&+&\frac{\gamma(D-2)(D-3)^2q^4}{2\al\xi(D-5)r^2},
\end{eqnarray}
due to both hypergeometric functions tend to one if $r\rightarrow +\infty$. The asymptotic relation (\ref{asymp_h_c5}) shows that for large radius the leading term is of anti-de Sitter (or de Sitter) type: {$\sim\frac{\al\xi}{(D-1)(D-2)}r^2$}, depending on the sign of the parameters $\al$ and $\xi$. {Here it is important to note} that this leading term does not depend on parameter $q$,  being completely defined by the parameters of the functions $G_2$ and $G_4$ {given previously in (\ref{case_E}) }. 

It should be also stressed that the representation (\ref{metr_hypergeo}) is valid only for even $D$ while for odd $D$ there are some subtleties, it is easy to see that the first hypergeometric function in (\ref{metr_hypergeo}) has some peculiarity if $D=5$ while the second one for $D=9$. But from the integral form (\ref{h_c5}) it follows that for odd $D$ the results of the integration might be written in terms of elementary functions. Namely, for $D=5$ we arrive at the following explicit form for the metric function $h(r)$ (if $\gamma>0$):
\begin{equation}
h(r)=-\frac{M}{r^2}+\frac{q^2}{4}\sqrt{\frac{\al^2\xi^2}{9q^4}r^4-\gamma}+\frac{15\gamma q^4}{4\al\xi r^2}\arcosh\left(\frac{\al\xi}{3q^2\sqrt{\gamma}}r^2\right),
\end{equation}
\textcolor{black}{whereas for} $\gamma<0$ instead of $\arcosh$ function one should utilize $\arsinh$. \textcolor{black}{For both cases, for a large $r$ we have that $$h(r) \simeq \frac{|\alpha \xi| r^{2}}{12}+O \left(\frac{1}{r^2}\right),$$
and we can find (A)dS planar configurations depending on the sign of the constants $\alpha$ and $\xi$ present in (\ref{case_E}) }

If $D=7$ the metric function $h(r)$ takes \textcolor{black}{an} even simpler form, namely it might be represented as a combination of irrational functions:
\begin{eqnarray}
\nonumber h(r)=-\frac{M}{r^4}+\frac{\al\xi}{30 r^4}\left(r^4-\frac{100\gamma q^4}{\al^2\xi^2}\right)^{\frac{3}{2}}\\+\frac{20\gamma q^4}{\al\xi r^4}\left(r^4-\frac{100\gamma q^4}{\al^2\xi^2}\right)^{\frac{1}{2}}.
\end{eqnarray}
It is straigthforward to check that for {large $r$}, the given exact relation gives rise to asymptotic relation of the form (\ref{asymp_h_c5}). Finally, \textcolor{black}{from the equation (\ref{G_22-1})}, the function ${\cal G}^2$ can be represented as:
\begin{equation}
{\cal G}^2=\frac{4 \xi^2}{ (1+\gamma X^2)^{\frac{3}{2}}},
\end{equation}
\textcolor{black}{while that the kinetic term $X$ takes the form found previously in (\ref{X_c5}) for seven dimensions}.
%%%%%%%%%%%%%%%%%%%%%%%%%%%%%%%%%%%%%
\subsection*{Case F}
%%%%%%%%%%%%%%%%%%%%%%%%%%%%%%%%%%%%%

As a final case to analyze,  we consider the following form for the functions $G_2$ and $G_4$:
\begin{equation}\label{case_F}
G_2=\al\left(1+\gamma X^2\right)^{7/4}, \quad G_4=\xi\left(1+\gamma X^2\right)^{1/4}.
\end{equation}
The equation (\ref{X_rel}) gives rise to:
\begin{equation}
X=-\frac{(D-2)(D-3)}{16\al\xi}\frac{q^2}{r^2},
\end{equation}
and the metric function $h(r)$ takes the form
\begin{eqnarray}\label{h_c6}
h(r)&=&-\frac{M}{r^{D-3}}+\frac{4\al\xi}{(D-1)(D-2)}r^2\nonumber\\
&+&\frac{3\gamma(D-2)(D-3)^2}{16 \al \xi (D-5)}\frac{q^4}{r^2}.
\end{eqnarray}
Here we would also like to note that if $D=5$ the latter term is of the form $\sim\ln(r)/r^2$, {where \textcolor{black}{for} this situation, as the Case D, the number  of horizons can be analyzed in terms of Lambert W functions \cite{BravoGaete:2019rci}}. The obtained metric function (\ref{h_c6}) is very similar to the asymptotic relation (\ref{asymp_h_c5}). In particular, the first and the second terms in these relations coincide completely, whereas the third terms have the same dependence on $r$, dimension of spacetime $D$, and the parameter $q$.

Finally, we can write the explicit form for the function ${\cal G}^2$ as:
\begin{equation}
{\cal G}^2=\xi^2\left(1+\frac{\gamma(D-2)^2(D-3)^2}{16\al^2\xi^2}\frac{q^4}{r^4}\right)^{-3/2}.
\end{equation}

%%%%%%%%%%%%%%%%%%%%%%%%%%%%%%%%%%%%%%%%%%%%%%%%%%%%%%%%%%%
\section{The time-independent case}\label{time-independent}
%%%%%%%%%%%%%%%%%%%%%%%%%%%%%%%%%%%%%%%%%%%%%%%%%%%%%%%%%%%

Now we will focus on a scalar field $\phi$ given by (\ref{phi-time}) with $q=0$, where the kinetic term $X$ is given from (\ref{X}), being  expressed as
\begin{eqnarray}
X=-\frac{1}{2}\,f\left(\varphi'\right)^2, \label{X-phi}
\end{eqnarray}
and the radial component $J^{r}=0$ of the equations of motions with respect to the scalar field (\ref{eq:jmu}) is satisfied imposing the condition (\ref{fh_time}).

\textcolor{black}{For the time-independent scalar field the equation (\ref{eq_Grr}),  the $(r,r)$-component  can be written in the following form:} 
\begin{equation}\label{difeqG2}
(r^{D-3}h)'\partial_{X}(G_2{\cal G})=0.
\end{equation}

{According to \cite{Kobayashi:2014eva}, the above equation allows us to conclude that we can find the kinetic term $X$ in an algebraic way, considering $X=$ constant such that $\partial_{X}
\left(G_{2}\,\mathcal{G}\right)=0$. Finally, the $(t,t)$-component can be expressed as a differential equation with respect to the metric function $f$ which reads
\begin{eqnarray}\label{difeqttd}
-{h}\,
\left[\frac{(D-2)\,\left(\mathcal{G}^{2}\,r^{D-3}\,f\right)'}
{\,r^{D-2}\,\mathcal{G}}-G_{2}\right]=0,
\end{eqnarray} 
\textcolor{black}{and the }expressions for the metric functions $f$ and $h$ \textcolor{black}{take the form}
\begin{eqnarray}
f(r)=h(r)=\frac{\,G_{2}}{(D-1)(D-2)\,\mathcal{G}}\,r^{2}-\frac{M}{r^{D-3}},\label{fh_sol}
\end{eqnarray}
where $M$, as before,  is an integration constant. \textcolor{black}{ For this situation, we can find a de-Sitter or Anti-de-Sitter config uration, where the radius takes the form:}
\begin{eqnarray}\label{eq:l}
\frac{\,G_{2}}{(D-1)(D-2)\,\mathcal{G}}:=l^{-2},
\end{eqnarray}
and the derivative of the scalar field as well as its explicit expression are given by
\begin{eqnarray}
&&(\phi')^{2}=-\frac{2\,X}{f},\nonumber\\
&&\phi(r)=\pm \frac{2\,\sqrt{-2 X l^{2}}\,
}{(D-1)}\nonumber\\
&&\times\ln\left[r^{\frac{D-3}{2}}\,\left(\frac{r}{l}
+\sqrt{\frac{r^2}{l^2}-\frac{M}{r^{D-3}}}\right)\right], \label{phi}
\end{eqnarray}
together with the condition
\begin{eqnarray*}
2 X l^{2} \leq 0,
\end{eqnarray*}
\textcolor{black}{while that the} remaining equations of motions with respect to the metric $\mathcal{E}_{x_i x_i}=0$ from (\ref{G_22-1}) are trivially satisfied.}

Here, it is worth pointing out that a particular form of the functions $G_2$ and $G_4$, when both of them are linear functions of $X$,  also belong to the class of functions for which $X=$constant is satisfied, where this \textcolor{black}{condition} stems directly from the \textcolor{black}{equations} of motion (\ref{emunu}) for the $(r,r)$- component. If we \textcolor{black}{consider the relation (\ref{case_1}), it follows from the equation (\ref{difeqG2}) that}
\begin{equation}
X=\frac{\b\xi-\al\gamma}{2\b\gamma}.
\end{equation}
Taking into account the equations  (\ref{fh_time}) (with $q=0$) and (\ref{difeqttd}),  we obtain:
\begin{equation}
f(r)=h(r)=-\frac{M}{r^{D-3}}+\frac{\beta}{(D-1)(D-2)\gamma}r^2.
\end{equation}

On the other hand, if we come back to the relation (\ref{difeqG2}) we can point out that this equation might be fulfilled if we do not impose the condition $X(r)=$constant as it has been performed above. In fact, from the condition $\partial_{X}(G_2{\cal G})=0$, \textcolor{black}{ we have that}
\begin{equation}
G_2{\cal G}=C_1,\label{g2G_cond}
\end{equation}
where $C_1$ is an integration constant. It is worth emphasizing that here we do not impose any condition on the kinetic term $X(r)$. The important conclusion which stems immediately from the relation (\ref{g2G_cond}) is the fact that the functions $G_2$ and $G_4$ in the Lagrangian (\ref{Lagrangian}) cannot be chosen independently, they have to obey the relation (\ref{g2G_cond}).

\textcolor{black}{From the equation} (\ref{difeqttd}) together with the condition (\ref{g2G_cond}), leads to the following relation:
\begin{equation}
f{\cal G}^2=-\frac{M}{r^{D-3}}+\frac{ C_1}{(D-1)(D-2)}r^2, \label{fG_prod}
\end{equation}
and here again, $M$ denotes an integration constant. We note that the right hand side of the relation (\ref{fG_prod}) is very similar to the metric functions $f$ and $h$ given by the relation (\ref{fh_sol}), but in the left hand side of the relation (\ref{fG_prod}) in contrast with (\ref{fh_sol}) we have product $f{\cal G}^2$. Surely, if the additional condition $X(r)=$constant is imposed there will be a complete coincidence of the results up to a redefinition of the constants. 

To obtain the metric function $h$, \textcolor{black}{we use} the relation (\ref{fh_time}) {with $q=0$}, once again taking into account the relations (\ref{g2G_cond}) and (\ref{fG_prod}), \textcolor{black}{we arrive at the following expression}
 \begin{equation}
h(r)=-\frac{M}{r^{D-3}}+\frac{C_1}{(D-1)(D-2)}r^2.\label{h_funct}
\end{equation}
\textcolor{black}{The above equation} (\ref{h_funct}) is of completely the same form as it is given by (\ref{fh_sol}). \textcolor{black}{Therefore,} the \textcolor{black}{explicit} form of metric function $h(r)$ (\ref{h_funct}) is valid even if  $X(r)=$constant is not imposed, but for a less restrictive condition, \textcolor{black}{given by} (\ref{g2G_cond}). At the same time, the equality $f(r)=h(r)$ \textcolor{black}{from} (\ref{fh_sol}) is violated in this a bit more general case. Here we would also like to stress that the form of the function ${\cal G}$ is completely defined by \textcolor{black}{a} chosen form of the functions $G_2$ or $G_4$. We might also expect that some constraints on the functions $G_2$ and $G_4$ might be imposed if one considers the field equation (\ref{emunu}) for the $(x_i,x_i)$- component, because it has not been used in the procedure described above. \textcolor{black}{Nevertheless}, a careful check shows that this equation does not impose any constraint, therefore the functions $G_2$ and $G_4$ might be chosen freely with the only condition, namely, they should obey (\ref{g2G_cond}).

%%%%%%%%%%%%%%%%%%
{\section{Thermodynamics for the time-independent solution}
\label{sec.termo}}
%%%%%%%%%%%%%%%%%%

\textcolor{black}{After} the steps performed to obtain the black hole solutions with planar base manifold {together with a time-independent scalar field}, in this section, we will explore their thermodynamic behavior by using the Wald formalism \cite{Wald:1993nt,Iyer:1994ys}, where the main idea \textcolor{black}{is the derivation of} the first law of black hole thermodynamics variating the Hamiltonian from a conserved Noether current. 

The variation of the action (\ref{action})-(\ref{Lagrangian}) reads
\begin{eqnarray*}
\textcolor{black}{\delta S=  \sqrt{-g}[ {\cal{E}}_{\mu \nu} \delta g^{\mu \nu} + {\cal{E}}_{\phi} \delta \phi]+\partial_{\mu} {\cal{J}}^{\mu},}
\end{eqnarray*}
where, as before,  \textcolor{black}{ ${\cal {E}}_{\mu \nu} $ and  ${\cal{E}}_{\phi} $ are the equations of motions with respect to the metric and the scalar field, given by (\ref{emunu})  and  (\ref{seom}) respectively, while ${\cal{J}}^{\mu}$ represents the surface term which reads}
\begin{eqnarray}
{\cal{J}}^{\mu}&=&\sqrt{-g}\Big[2\left(P^{\mu (\alpha\beta)
\gamma}\nabla_{\gamma}\delta g_{\alpha\beta}-\delta g_{\alpha\beta} \nabla_{\gamma}P^{\mu(\alpha\beta)\gamma}\right) \nonumber \\
&+&\frac{\delta \cal{L}}{\delta (\phi_{\mu})} \delta
\phi-\nabla_{\nu}\left(\frac{\delta \cal{L}}{\delta (\phi_{\mu
\nu})}\right) \delta \phi
+\frac{\delta \cal{L}}{\delta (\phi_{\mu \nu})} \delta (\phi_{\nu})\nonumber \\
&-&\frac{1}{2}\frac{\delta \cal{L}}{\delta (\phi_{\mu \rho})}
\phi^{\sigma} \,
\delta g_{\sigma \rho}-\frac{1}{2}\frac{\delta \cal{L}}{\delta (\phi_{\rho \mu})}\phi^{\sigma} \,\delta g_{\sigma \rho} \nonumber \\
&+&\frac{1}{2}\frac{\delta \cal{L}}{\delta ( \phi_{\sigma
\rho})}\phi^{\mu}\,\delta g_{\sigma \rho}\Big],\label{eq:surface}
\end{eqnarray}
where we have:
\begin{eqnarray*}
P^{\mu\nu\lambda\rho}&=&\frac{\delta {\cal{L}}}{ \delta R_{\mu\nu\lambda\rho}}=\frac{1}{2} G_{4} \left(g^{\mu \lambda} g^{\nu \rho}-g^{\mu \rho} g^{\nu \lambda}\right),\\
\frac{\delta \cal{L}}{\delta (\phi_{\mu})}&=&J^{\mu},\qquad  \frac{\delta \cal{L}}{\delta ( \phi_{\sigma
\rho})}= 2 G_{4X} (\Box \phi g^{\sigma \rho}-\nabla^{\sigma} \nabla^{\rho} \phi),
\end{eqnarray*}
\textcolor{black}{with} $J^{\mu}$ given previously in (\ref{eq:jmu}).

In particular, \textcolor{black}{for the metric (\ref{metricd}) and a time independent scalar field $\phi=\phi(r)$,} the only nonzero component of the surface current ${\cal{J}}^{\mu}$ is the radial component ${\cal{J}}^{r}$ which takes the following form:
\begin{eqnarray}
&&{\cal{J}}^{r}=\sqrt{\frac{h}{f}}r^{D-2}\left\{-G_4\left[\frac{f}{h}\left(\delta h'+\frac{h'}{2}\left(\frac{\delta f}{f}-\frac{\delta h}{h}\right)\right)\right.\right.\nonumber\\
&&\left.\left.\nonumber+\frac{D-2}{r}\delta f\right]+G_{4X}f(\vp')^2\left[\left(\frac{h'}{2h}+\frac{D-2}{r}\right)\delta f+\right.\right.\\
&&\left.\left.\frac{(D-2)f}{rh}\delta h\right]+2G_{4X}f^2\left(\frac{h'}{2h}+\frac{D-2}{r}\right)\vp'\delta\vp'\right\}.\label{surf:current}
\end{eqnarray}
It is worth noting that the current $J^{r}$ \textcolor{black}{from (\ref{eq:jmu}) vanishes}, therefore the term proportional to $\delta\phi$ does not appear in the relation (\ref{surf:current})

{To compute the entropy, by using the surface term given in (\ref{eq:surface}) we define a $1$-form ${\cal{J}}_{(1)}={\cal{J}}_{\mu} dx^{\mu}$ and its Hodge dual ${\Theta}_{(D-1)}=(-1)*{\cal{J}}_{(1)}$. Then, after making use of the equations of motion (this is ${\cal{E}}_{\mu \nu}=0$ and ${\cal{E}}_{\phi}=0$), we have 
$${\cal{J}}_{(D-1)}={\Theta}_{(D-1)}-i_{\xi}*\mathcal{L}=-d* {\cal{J}}_{(2)},$$
where $i_{\xi}$ is a contraction of the vector field $\xi^{\mu}$ on the first index of $*\mathcal{L}$. The above relation allows to define a $(D-2)$-form ${Q}_{(D-2)}=*{\cal{J}}_{(2)}$ such that ${\cal{J}}_{(D-1)}=d Q_{(D-2)}$, where
\textcolor{black}{$$Q_{(D-2)}=Q_{\alpha_1 \alpha_2 \cdots \alpha_{D-2}}=\epsilon_{\alpha_1 \alpha_2 \cdots \alpha_{D-2} \mu \nu} Q^{\mu \nu}$$
with
\begin{eqnarray}
Q^{\mu\nu}&=&\Big[2P^{\mu\nu\rho\sigma}\nabla_\rho \xi_\sigma -4\xi_\sigma
\nabla_\rho P^{\mu\nu\rho\sigma}\nonumber \\
&+&\frac{\delta \cal{L}}{ \delta \phi_{\mu \sigma}} \phi^{\nu} \xi_{\sigma}-\frac{\delta \cal{L}}{ \delta \phi_{\nu \sigma}} \phi^{\mu}
\xi_{\sigma}\Big].\label{eq:noether}
\end{eqnarray}
To obtain} the first law of black hole thermodynamics, the vector field $\xi^{\mu}$ is supposed to be a time-translation vector, which is \textcolor{black}{a} Killing vector and it is null on the event horizon $r_h$. In particular, \textcolor{black}{for a planar black hole metric,} we obtain
\begin{equation}
Q^{tr}=\sqrt{\frac{h}{f}}r^{D-2}\left(-\frac{f}{h}G_{4}h'+\frac{2(D-2)}{r}G_{4X}f^2(\vp')^2\right).\label{noet:tr}
\end{equation}
Finally, the variation of the Hamiltonian reads
\begin{eqnarray}
\delta \mathcal{H}&=&\delta \int_{\mathcal{C}} {\cal{J}}_{(D-1)} -\int_{\mathcal{C}} d \left(i_{\xi} \Theta_{(D-1)}\right) \nonumber \\
&=& \int_{\Sigma^{(D-2)}}\left(\delta {Q}_{(D-2)}-i_{\xi} {\Theta}_{(D-1)}\right),\nonumber \\
&=&- \Big\{(D-2)\sqrt{\frac{h}{f}}r^{D-3}\Big[G_{4}\delta f \nonumber \\
&+&\ 2G_{4X}f \left(\frac{(\vp')^2}{2}  \delta f-\delta X\right) \nonumber \\
&-&4G_{4XX}fX\delta X\Big]\Sigma_{D-2}\Big\}\nonumber\\
&=&-\frac{(D-2)}{ {\cal{G}}} \sqrt{\frac{h}{f}} r^{D-3}  \Sigma_{D-2} \delta  ({\cal{G}}^2 f),\label{var_diff}
\end{eqnarray}
where $\mathcal{C}$ and $\Sigma^{(D-2)}$ are a Cauchy Surface and its boundary respectively,  $ \Sigma_{D-2} $ is the finite volume of the $(D-2)$-dimensional compact angular base manifold, ${\cal{G}}$ was given previously in (\ref{functionG}) and $\delta X$ denotes variation of the kinetic term for the scalar field $X$, namely, it equals to $\delta X=-\left(\frac{(\vp')^2}{2}\delta f+f\vp'\delta\vp'\right)$. Here we note that (\ref{var_diff}) has two components, one of them located at infinity, and denoted as $ \mathcal{H}_{\infty}$, and the other at the horizon, given by $\mathcal{H}_{+}$, and due to the fact that there are two types of  black hole solutions for a time-independent scalar field, we split the following analysis into two branches, namely one of them is for the configuration (\ref{fh_sol})-(\ref{phi}), we find
\begin{eqnarray}
\delta {\mathcal{H}}_{\infty}&=& (D-2){\cal{G}} \Sigma_{D-2} \delta M , \nonumber\\
\delta {\mathcal{H}}_{+}&=& \frac{(D-2)(D-1){\cal{G}} \Sigma_{D-2} r_h^{D-2}}{ l^2} \delta r_h  \nonumber\\
&=& T \delta \left( 4 \pi r_h^{D-2}  {\cal{G}} \Sigma_{D-2}\right) ,  \label{dH}
\end{eqnarray}
with the \textcolor{black}{(A)dS}-radius $l$ given previously in (\ref{eq:l}), and $T$ is the Hawking Temperature
\begin{eqnarray}
T=\frac{\kappa}{2 \pi}\Big{|}_{r=r_h}=\frac{1}{4\pi} \left(h'(r) \sqrt{\frac{f(r)}{h(r)}}\right) \Big{|}_{r=r_h}, \label{eq:thgen}
\end{eqnarray}
constructed by the surface gravity $\kappa$ \textcolor{black}{which reads}
\begin{equation}
\kappa=\sqrt{-\frac{1}{2}\left(\nabla_{\mu} \xi_{\nu}\right)\left(\nabla^{\mu} \xi^{\nu}\right)},\label{bh_T}
\end{equation}
\textcolor{black}{and} the time-like Killing vector $\partial_{t}=\xi^{\mu}\partial_{\mu}$, and given by
\begin{equation}
T= \frac{(D-1)r_h}{4 \pi l^{2}}.\label{th}
\end{equation}
According to Wald procedure, the equality $\delta \mathcal{H}_{\infty}=\delta \mathcal{H}_{+}$ implies the first law of black holes thermodynamics
\begin{equation}
d\mathcal{M}= T d\mathcal{S}_{W},\label{eq:first-law}
\end{equation}
where from equations (\ref{dH}) 
\begin{eqnarray*}
{\mathcal{M}}&=&(D-2){\cal{G}} \Sigma_{D-2} M =\frac{(D-2){{\cal{G}} r_h^{D-1}} \Sigma_{D-2}}{ l^{2}} \label{masssol1},\\
{\mathcal{S}}_{W}&=& 4 \pi   {\cal{G}} r_h^{D-2} \Sigma_{D-2}.
\end{eqnarray*}
\textcolor{black}{In order to have positive extensive thermodynamical quantities, we consider the case of AdS-planar black holes, where its radius $l$ takes the form  (\ref{eq:l}) with ${\,G_{2}}/{\,\mathcal{G}}>0$}.  Just for completeness, a higher dimensional Smarr relation \cite{Smarr:1972kt}
\begin{equation}\label{eq:smarr}
\mathcal{M}= \left(\frac{D-2}{D-1}\right)\,T \mathcal{S}_{W},   
\end{equation}
 is satisfied.
 
On the other hand, and following the same steps as before, for the solution  (\ref{g2G_cond})-(\ref{h_funct}) we have
\begin{eqnarray*}
\delta\mathcal{H}_{\infty}&=&(D-2)\delta M\Sigma_{D-2}=\delta\mathcal{M}, \\
&\Rightarrow& {\mathcal{M}}=(D-2) \Sigma_{D-2} M =\frac{ C_1 r_h^{D-1} \Sigma_{D-2}}{(D-1)} \label{masssol2},\\
T&=&\frac{C_1 r_h}{ 2 \pi (D-2) {\cal{G}}(X(r_h))},
\end{eqnarray*}
where we suppose that 
${\cal{G}} $ is a positive function for $r \geq r_h$, \textcolor{black}{ while that $C_1$ and $M$ are positive integration constants}. Together with the above, to satisfy the first law (\ref{eq:first-law})
\begin{eqnarray*}
d{\mathcal{M}}&=&{ C_1 r_h^{D-2} \Sigma_{D-2} d r_h}=T d {\mathcal{S}}_{W} \\
&=& \frac{C_1 r_h}{ 4 \pi (D-2) {\cal{G}} (X(r_h))} d {\mathcal{S}}_{W}\\
&\Rightarrow& {\mathcal{S}}_{W}= 4\pi (D-2) \int d r_h  {\cal{G}} (X(r_h)) r_h^{D-3} \Sigma_{D-2}.
\end{eqnarray*} 
Obviously, if ${\cal{G}} (X(r_h))$ is a constant that \textcolor{black}{does} not depends on the location of the event horizon $r_h$, \textcolor{black}{we have that}
$${\mathcal{S}}_{W}= 4\pi  {\cal{G}} r_h^{D-2} \Sigma_{D-2},$$ fulfilling, in addition, the Smarr relation (\ref{eq:smarr}).

Here it is important to note that for some cases the \textcolor{black}{Wald's relation} for the entropy (or more generally \textcolor{black}{the Wald's approach}) might be ambiguous, this ambiguity appears in particular for Horndeski-type theories. To cure these ambiguities, the so-called solution phase-space method (SPSM) was established, which can be treated as a further elaboration of Wald's approach \cite{Hajian:PRD2016}. \textcolor{black}{A recent} study also shows that the definition of the black hole temperature should be modified for \textcolor{black}{the} Horndeski theory, \textcolor{black}{and it can be explained due to} different speeds of propagations of photons and gravitons  \cite{Hajian:PLB2020}. Therefore, the \textcolor{black}{black holes's} temperature now can be defined as follows
\begin{equation}
T_{bh}={\cal G}T,\label{T_bh_m}
\end{equation}
where $T$ is the Hawking temperature (\ref{eq:thgen}) which is completely defined by the surface gravity $\kappa$ (\ref{bh_T}). Now, from the relation at the infinity $\delta \mathcal{H}_{\infty} $ for the time-independent scalar field solutions  (\ref{fh_sol})-(\ref{phi}) and  (\ref{g2G_cond})-(\ref{h_funct}), we can infer that
$${\mathcal{M}}=(D-2)\Sigma_{D-2} M,$$
where we point out that here $\mathcal{M}$ might be treated as a mass (thermodynamic) of the black hole.

On the other hand, considering the relation  $\delta \mathcal{H}_{+} $  and taking into account the relation (\ref{T_bh_m}), we can write:
\begin{equation}
\nonumber\delta\mathcal{H}_{+}=4\pi(D-2)T_{bh}r^{D-3}_{+}\delta r_{h}\Sigma_{D-2}=T_{bh}\delta\mathcal{\bar{S}}_{W},
\end{equation}
where now $\mathcal{\bar{S}}_{W}$ is the entropy, which takes the following form
\begin{equation}
\mathcal{\bar{S}}_{W}=4\pi\Sigma_{D-2}r^{D-2}_{h}.
\end{equation}
Here we have \textcolor{black}{an agreement} with the corresponding relation in the framework of standard General Relativity. Now we are able to write the first law of black \textcolor{black}{hole} thermodynamics, which now takes the form:
$$
d\mathcal{M}=T_{bh}d\mathcal{\bar{S}}_{W}.
$$
It can be easily shown that a Smarr relation
$$
\mathcal{M}=\left(\frac{D-2}{D-1}\right)T_{bh}\mathcal{\bar{S}}_W,
$$
also holds.\\

%%%%%%%%%%%%%%%%%%
\section{Exploring the {viscosity/entropy density}  ratio}
\label{viscosity}
%%%%%%%%%%%%%%%%%%

As was \textcolor{black}{mentioned} in the introduction, \textcolor{black}{ planar black hole configurations have a particularity which allows us} to study the viscosity/\textcolor{black}{entropy density}  $(\eta/s)$ ratio, wherein our case are the \textcolor{black}{AdS} solutions \textcolor{black}{given by }(\ref{fh_sol})-(\ref{phi}). As a first step, we perform a transverse and traceless perturbation \textcolor{black}{of} the metric (\ref{metricd}) for $D >3$ with $h=f$, which reads 
\begin{equation}
ds^2=-f(r) dt^2+\frac{dr^2}{f(r)}+2 r^2 \Psi(t,r) dx_1 dx_2+ r^2 \sum_{i=1}^{D-2} d x_{i}^2,
\end{equation}
 where \textcolor{black}{for} the Ansatz 
$$\Psi(t,r) =\zeta t + h_{x_1 x_2}(r),$$
with $\zeta$ a constant identified as the gradient of the fluid velocity along the $x_1$ direction \cite{Fan:2018qnt}, yields the following linearized equations for $h_{x_1 x_2}$:
\begin{eqnarray}\label{linear}
\left[{{\cal{G}}} r^{D-2} f (h_{x_1 x_2})'\right]'=0.
\end{eqnarray}
According to the Wald formalism \cite{Iyer:1994ys,Wald:1993nt} together with the method \cite{Fan:2018qnt}, the shear stress is associated \textcolor{black}{with} the current
\begin{equation}\label{eq:current}
{\mathcal{J}}^{x_2}= \sqrt{-g} {Q}^{r x_2}= {\cal{G}}\,r^{D-2} f (h_{x_1 x_2})', 
\end{equation}\\
\textcolor{black}{where $Q^{r x_2}$ is defined by (\ref{eq:noether}) together with a space-like Killing vector $\partial_{x_1}=\xi^{\mu}\partial_{\mu}$. The current (\ref{eq:current}) is conserved due to the linear equation (\ref{linear}).} Imposing the ingoing horizon boundary condition
$$h_{x_1 x_2} = \zeta  \sqrt{\frac{ G_4}{{\cal{G}}}}\,\frac{\log (r-r_h)}{4 \pi T}+\cdots,$$\\
as well as a Taylor expansion in the near horizon region $r_h$, \textcolor{black}{that} is:
$$h=f=4\pi T (r-r_h)+\cdots,$$
where $T$ is the Hawking temperature (\ref{th}), we have:
$$ \eta = \zeta  {\cal{G}}\, \sqrt{\frac{G_4}{{\cal{G}}}}\, r_h^{D-2}=   \frac{1}{4 \pi}  \sqrt{\frac{ G_4}{{\cal{G}}}} \zeta s,  $$
with the entropy density $s$ given by
$$s=\frac{\mathcal{S}_{W}}{\Sigma_{D-2}}={4 \pi r_h^{D-2} {\cal{G}}}.$$
\textcolor{black}{Finally, the viscosity/ entropy density ratio  takes the form:
\begin{equation}\label{vis/den}
\frac{\eta}{s}=\frac{1}{4 \pi}  \sqrt{\frac{G_4}{{\cal{G}}}}=\frac{1}{4 \pi}  \sqrt{\frac{G_4}{{G_4-2 X G_{4X}}}}.
\end{equation}
 \\
Some comments about the above result can be made. First that all, similarly to the cases \cite{Buchel:2003tz,Buchel:2004qq,Benincasa:2006fu,Landsteiner:2007bd}, the  ($\eta/s$)  ratio (\ref{vis/den}) does not depend on the event horizon $r_h$. Nevertheless, together with the constant $1/(4\pi)$ a contribution depending on $G_4$ and ${\mathcal{G}}$ appears}.  In addition, \textcolor{black}{the  viscosity/ entropy density ratio} for the linear case (\ref{case_1}), analyzed in \cite{Feng:2015oea}, can be recovered.  As the linear situation,  the above result allows us to construct examples where
\textcolor{black}{$$0<\frac{G_4}{{\cal{G}}}=\frac{G_4}{{G_4-2 X G_{4X}}}<1,$$
with a suitable choice of parameters, violating the KSS bound,  for example, for $X G_{4X}<0$ and $G_4>0$. It is worth pointing out that for the situation $G_{4}=$constant, we have the saturated situation (that is $\eta/s=1/(4\pi)$).} Just for completness, the viscosity/ \textcolor{black}{entropy density}  $(\eta/s)$ ratio (\ref{vis/den}) also can be found following the steps described in \cite{Son:2002sd,Brigante:2007nu}.

%%%%%%%%%%%%%%%%%%
\section{Conclusions and Discussions}
\label{conclusions}
%%%%%%%%%%%%%%%%%%

In the present paper, we explore new hairy black hole solutions in arbitrary dimensions and a planar base manifold based on the work developed in \cite{Kobayashi:2014eva}. \textcolor{black}{For} this case, the model is given by a special case of the Hordeski theory with shift symmetry and reflection symmetry  (\ref{action})-(\ref{Lagrangian}), 
constructed by two coupling functions depending on the kinetic term $X$.  For these configurations, we split our analysis based on the time-dependence or time independence of the scalar field, obtaining a set of new configurations, depending on the relation between the coupling functions $G_2$ and $G_4$.  \textcolor{black}{It is worth pointing out that depending on the form of the function $\cal{G}$ (\ref{functionG}), constructed through the kinetic term $X$ and the function $G_4$, together with the equation (\ref{int_G_tt}), we can find planar black holes configurations with various asymptotic behaviors.  For instance, there are asymptotically (A)dS solutions (given by the cases C to F) for linear time-dependent scalar field configurations as well as the for the time-independent case. On the other side, there are Lifshitz-type solutions and other generalizations (obtained in the cases A and B).}

Together with the above, the thermodynamics is analyzed for the time-independent situation through the Wald formalism \cite{Wald:1993nt,Iyer:1994ys} . It is worth pointing out that the thermodynamical quantities are not unique, \textcolor{black}{due to} the ambiguity present in the Wald approach, \textcolor{black}{which becomes remarkable for} this kind of scalar-tensor theory. To circumvent this inconvenience, we consider the solution phase space method (SPSM), which can be treated as a further elaboration of Wald's method \cite{Hajian:PRD2016},  redefining the black hole's temperature  \cite{Hajian:PLB2020} and implying that the mass and the entropy resemble the standard General Relativity extensive quantities. For both situations, the First Law, as well as a higher dimensional Smarr relation, are fulfilled.

In addition, the Wald procedure  \cite{Wald:1993nt,Iyer:1994ys} together with the method \cite{Fan:2018qnt} allows us to compute the shear viscosity/ \textcolor{black}{entropy density}  $(\eta/s)$ ratio, wherein our situation depends on the coupling functions $G_2$ and $G_4$, showing a new \textcolor{black}{specific} example where the KSS bound  (\ref{KSS}) can be violated.

Some natural extensions of this work would be for example to consider spherical or hyperbolical topologies for the event horizon, the inclusions of electromagnetic sources (see for example \cite{BravoGaete:2019rci,Plebanski:1968,Alvarez:2014pra,Stetsko:2020nxb,Stetsko:2020tjg} ), or even a recently extension denominated as
Degenerate-Higher-Order-Scalar-Tensor (DHOST) theory, allowing us to add new degrees of freedom introducing a scalar field, avoiding Ostrogradsky instability due to its degeneracy property  \textcolor{black}{\cite{BenAchour:2016fzp,Motohashi:2016ftl,Babichev:2020qpr,Baake:2020tgk,Baake:2021kyg,Baake:2021jzv,Santos:2020lmb}}}.

%%%%%%%%%%%%%%%%%
\begin{acknowledgments}
%%%%%%%%%%%%%%%%%

{MB would like to thank L. Guajardo for useful discussions and comments on this work. The authors thank the Referee for the
commentaries and suggestions to improve the paper.}

\end{acknowledgments}

%%%%%%%%%%%%%%%%%%%%%%%%%%%%%%%%%%%%%%%%%%%%%%%%%%%%%%%%%
\appendix
\section{Equations of motions with respect to the metric}
%%%%%%%%%%%%%%%%%%%%%%%%%%%%%%%%%%%%%%%%%%%%%%%%%%%%%%%%%
In the following section, we present the equations of motions 
 (\ref{emunu}) based on the computations performed in \cite{Kobayashi:2011nu}
\begin{widetext}
\begin{eqnarray*}
{\cal G}_{\mu
\nu}^1&=&-\frac{1}{2}G_{2X}\nabla_\mu\phi\nabla_\nu\phi-\frac{1}{2}\,
G_{2}\,g_{\mu\nu},
\\
{\cal G}_{\mu \nu}^2&=&
G_4G_{\mu\nu}-\frac{1}{2}G_{4X}R\nabla_\mu\phi\nabla_\nu\phi
-\frac{1}{2}G_{4XX}\left[(\Box\phi)^2-(\nabla_\alpha\nabla_\beta\phi)^2\right]\nabla_\mu
\phi\nabla_\nu\phi \cr&& -G_{4X}\Box\phi\nabla_\mu\nabla_\nu\phi
+G_{4X}\nabla_\lambda\nabla_\mu\phi\nabla^\lambda\nabla_\nu\phi
+2\nabla_\lambda
G_{4X}\nabla^\lambda\nabla_{(\mu}\phi\nabla_{\nu)}\phi-\nabla_\lambda
G_{4X}\nabla^\lambda\phi\nabla_\mu\nabla_\nu\phi \cr&&
+g_{\mu\nu}\biggl\{
G_{4XX}\nabla_\alpha\nabla_\lambda\phi\nabla_\beta\nabla^\lambda\phi\nabla^\alpha\phi\nabla^\beta\phi
+\frac{1}{2}G_{4X}\left[(\Box\phi)^2-(\nabla_\alpha\nabla_\beta\phi)^2\right]\biggr\}
\cr&&
+2\Bigl[G_{4X}R_{\lambda(\mu}\nabla_{\nu)}\phi\nabla^\lambda\phi-\nabla_{(\mu}G_{4X}\nabla_{\nu)}\phi\Box\phi\Bigr]
-g_{\mu\nu}\left[
G_{4X}R^{\alpha\beta}\nabla_\alpha\phi\nabla_\beta\phi-\nabla_\lambda
G_{4X}\nabla^\lambda\phi \Box\phi\right] \cr&&
+G_{4X}R_{\mu\alpha\nu\beta}\nabla^\alpha\phi\nabla^\beta\phi
-G_{4XX}\nabla^\alpha\phi\nabla_\alpha\nabla_\mu\phi
\nabla^\beta\phi\nabla_\beta\nabla_\nu\phi,
\end{eqnarray*}
\end{widetext}
where the equations are given by
\begin{eqnarray*}
{\cal E}_{\mu\nu}=\sum_{i=1}^2{\cal G}_{\mu\nu}^i=0.
\end{eqnarray*}

%%%%%%%%%%%%%%%%%%%%%%%%%%%


\begin{thebibliography}{99}
%%%%%%%%%%%%%%%%%%%%%%%%%%%

%\cite{Riess:1998cb}
\bibitem{Riess:1998cb}
A.~G.~Riess \textit{et al.} [Supernova Search Team],
%``Observational evidence from supernovae for an accelerating universe and a cosmological constant,''
Astron. J. \textbf{116} (1998), 1009-1038
doi:10.1086/300499
[arXiv:astro-ph/9805201 [astro-ph]].
%12922 citations counted in INSPIRE as of 04 Dec 2020

%\cite{Perlmutter:1998np}
\bibitem{Perlmutter:1998np}
S.~Perlmutter \textit{et al.} [Supernova Cosmology Project],
%``Measurements of $\Omega$ and $\Lambda$ from 42 high redshift supernovae,''
Astrophys. J. \textbf{517} (1999), 565-586
doi:10.1086/307221
[arXiv:astro-ph/9812133 [astro-ph]].
%12953 citations counted in INSPIRE as of 04 Dec 2020

%\cite{Abbott:2016blz}
\bibitem{Abbott:2016blz}
B.~P.~Abbott \textit{et al.} [LIGO Scientific and Virgo],
%``Observation of Gravitational Waves from a Binary Black Hole Merger,''
Phys. Rev. Lett. \textbf{116} (2016) no.6, 061102
doi:10.1103/PhysRevLett.116.061102
[arXiv:1602.03837 [gr-qc]].
%5864 citations counted in INSPIRE as of 04 Dec 2020

%\cite{Horndeski:1974wa}
\bibitem{Horndeski:1974wa}
  G.~W.~Horndeski,
  %``Second-order scalar-tensor field equations in a four-dimensional space,''
  Int.\ J.\ Theor.\ Phys.\  {\bf 10}, 363 (1974).
  %%CITATION = IJTPB,10,363;%%
  %147 citations counted in INSPIRE as of 26 Dec 2013



%\cite{Deffayet:2009mn}
\bibitem{Deffayet:2009mn}
C.~Deffayet, S.~Deser and G.~Esposito-Farese,
%``Generalized Galileons: All scalar models whose curved background extensions maintain second-order field equations and stress-tensors,''
Phys. Rev. D \textbf{80} (2009), 064015
doi:10.1103/PhysRevD.80.064015
[arXiv:0906.1967 [gr-qc]].
%592 citations counted in INSPIRE as of 04 Dec 2020

%\cite{Charmousis:2014mia}
\bibitem{Charmousis:2014mia}
C.~Charmousis,
%``From Lovelock to Horndeski`s Generalized Scalar Tensor Theory,''
Lect. Notes Phys. \textbf{892} (2015), 25-56
doi:10.1007/978-3-319-10070-8\_2
[arXiv:1405.1612 [gr-qc]].
%50 citations counted in INSPIRE as of 04 Dec 2020


%\cite{Kobayashi:2011nu}
\bibitem{Kobayashi:2011nu}
  T.~Kobayashi, M.~Yamaguchi and J.~'i.~Yokoyama,
  %``Generalized G-inflation: Inflation with the most general second-order field equations,''
  Prog.\ Theor.\ Phys.\  {\bf 126} (2011) 511
  [arXiv:1105.5723 [hep-th]].
  %%CITATION = ARXIV:1105.5723;%%
  %124 citations counted in INSPIRE as of 25 Apr 2014



\bibitem{Maldacena:1997re}
J.~M.~Maldacena,
%``The Large N limit of superconformal field theories and supergravity,''
Adv. Theor. Math. Phys. \textbf{2} (1998), 231-252
doi:10.1023/A:1026654312961
[arXiv:hep-th/9711200 [hep-th]].
%16594 citations counted in INSPIRE as of 30 Apr 2021





\bibitem{Gubser:1998bc}
S.~S.~Gubser, I.~R.~Klebanov and A.~M.~Polyakov,
%``Gauge theory correlators from noncritical string theory,''
Phys. Lett. B \textbf{428} (1998), 105-114
doi:10.1016/S0370-2693(98)00377-3
[arXiv:hep-th/9802109 [hep-th]].
%9058 citations counted in INSPIRE as of 30 Apr 2021



\bibitem{Witten:1998qj}
E.~Witten,
%``Anti-de Sitter space and holography,''
Adv. Theor. Math. Phys. \textbf{2} (1998), 253-291
doi:10.4310/ATMP.1998.v2.n2.a2
[arXiv:hep-th/9802150 [hep-th]].
%10681 citations counted in INSPIRE as of 30 Apr 2021





\bibitem{Policastro:2001yc}
G.~Policastro, D.~T.~Son and A.~O.~Starinets,
%``The Shear viscosity of strongly coupled N=4 supersymmetric Yang-Mills plasma,''
Phys. Rev. Lett. \textbf{87} (2001), 081601
doi:10.1103/PhysRevLett.87.081601
[arXiv:hep-th/0104066 [hep-th]].
%1454 citations counted in INSPIRE as of 30 Apr 2021



\bibitem{Son:2002sd}
D.~T.~Son and A.~O.~Starinets,
%``Minkowski space correlators in AdS / CFT correspondence: Recipe and applications,''
JHEP \textbf{09} (2002), 042
doi:10.1088/1126-6708/2002/09/042
[arXiv:hep-th/0205051 [hep-th]].
%1035 citations counted in INSPIRE as of 30 Apr 2021



\bibitem{Kovtun:2003wp}
P.~Kovtun, D.~T.~Son and A.~O.~Starinets,
%``Holography and hydrodynamics: Diffusion on stretched horizons,''
JHEP \textbf{10} (2003), 064
doi:10.1088/1126-6708/2003/10/064
[arXiv:hep-th/0309213 [hep-th]].
%627 citations counted in INSPIRE as of 30 Apr 2021


\bibitem{Kovtun:2004de}
P.~Kovtun, D.~T.~Son and A.~O.~Starinets,
%``Viscosity in strongly interacting quantum field theories from black hole physics,''
Phys. Rev. Lett. \textbf{94} (2005), 111601
doi:10.1103/PhysRevLett.94.111601
[arXiv:hep-th/0405231 [hep-th]].
%2464 citations counted in INSPIRE as of 30 Apr 2021






\bibitem{Buchel:2003tz}
A.~Buchel and J.~T.~Liu,
%``Universality of the shear viscosity in supergravity,''
Phys. Rev. Lett. \textbf{93} (2004), 090602
doi:10.1103/PhysRevLett.93.090602
[arXiv:hep-th/0311175 [hep-th]].
%493 citations counted in INSPIRE as of 30 Apr 2021




\bibitem{Buchel:2004qq}
A.~Buchel,
%``On universality of stress-energy tensor correlation functions in supergravity,''
Phys. Lett. B \textbf{609} (2005), 392-401
doi:10.1016/j.physletb.2005.01.052
[arXiv:hep-th/0408095 [hep-th]].
%171 citations counted in INSPIRE as of 30 Apr 2021



\bibitem{Benincasa:2006fu}
P.~Benincasa, A.~Buchel and R.~Naryshkin,
%``The Shear viscosity of gauge theory plasma with chemical potentials,''
Phys. Lett. B \textbf{645} (2007), 309-313
doi:10.1016/j.physletb.2006.12.030
[arXiv:hep-th/0610145 [hep-th]].
%94 citations counted in INSPIRE as of 30 Apr 2021



\bibitem{Landsteiner:2007bd}
K.~Landsteiner and J.~Mas,
%``The Shear viscosity of the non-commutative plasma,''
JHEP \textbf{07} (2007), 088
doi:10.1088/1126-6708/2007/07/088
[arXiv:0706.0411 [hep-th]].
%62 citations counted in INSPIRE as of 30 Apr 2021




\bibitem{Iqbal:2008by}
N.~Iqbal and H.~Liu,
%``Universality of the hydrodynamic limit in AdS/CFT and the membrane paradigm,''
Phys. Rev. D \textbf{79} (2009), 025023
doi:10.1103/PhysRevD.79.025023
[arXiv:0809.3808 [hep-th]].
%495 citations counted in INSPIRE as of 30 Apr 2021



\bibitem{Cai:2008ph}
R.~G.~Cai, Z.~Y.~Nie and Y.~W.~Sun,
%``Shear Viscosity from Effective Couplings of Gravitons,''
Phys. Rev. D \textbf{78} (2008), 126007
doi:10.1103/PhysRevD.78.126007
[arXiv:0811.1665 [hep-th]].
%142 citations counted in INSPIRE as of 30 Apr 2021



\bibitem{Cai:2009zv}
R.~G.~Cai, Z.~Y.~Nie, N.~Ohta and Y.~W.~Sun,
%``Shear Viscosity from Gauss-Bonnet Gravity with a Dilaton Coupling,''
Phys. Rev. D \textbf{79} (2009), 066004
doi:10.1103/PhysRevD.79.066004
[arXiv:0901.1421 [hep-th]].
%132 citations counted in INSPIRE as of 30 Apr 2021


\bibitem{Wald:1993nt}
  R.~M.~Wald,
  {\it Black hole entropy is the Noether charge},
  Phys.\ Rev.\ D {\bf 48}, no. 8, R3427 (1993),
  [gr-qc/9307038].
  %%CITATION = doi:10.1103/PhysRevD.48.R3427;%%
  %1548 citations counted in INSPIRE as of 11 Nov 2019


\bibitem{Iyer:1994ys}
  V.~Iyer and R.~M.~Wald,
  {\it Some properties of Noether charge and a proposal for dynamical black hole entropy},
  Phys.\ Rev.\ D {\bf 50}, 846 (1994),
  [gr-qc/9403028].
  %%CITATION = doi:10.1103/PhysRevD.50.846;%%
  %1310 citations counted in INSPIRE as of 11 Nov 2019



%\cite{Fan:2018qnt}
\bibitem{Fan:2018qnt}
Z.~Y.~Fan,
%``Note on the Noether charge and holographic transports,''
Phys. Rev. D \textbf{97} (2018) no.6, 066013
doi:10.1103/PhysRevD.97.066013
[arXiv:1801.07870 [hep-th]].
%10 citations counted in INSPIRE as of 30 Apr 2021



\bibitem{Kats:2007mq}
Y.~Kats and P.~Petrov,
%``Effect of curvature squared corrections in AdS on the viscosity of the dual gauge theory,''
JHEP \textbf{01} (2009), 044
doi:10.1088/1126-6708/2009/01/044
[arXiv:0712.0743 [hep-th]].
%354 citations counted in INSPIRE as of 30 Apr 2021




\bibitem{Brigante:2007nu}
M.~Brigante, H.~Liu, R.~C.~Myers, S.~Shenker and S.~Yaida,
%``Viscosity Bound Violation in Higher Derivative Gravity,''
Phys. Rev. D \textbf{77} (2008), 126006
doi:10.1103/PhysRevD.77.126006
[arXiv:0712.0805 [hep-th]].
%522 citations counted in INSPIRE as of 30 Apr 2021






%\cite{Feng:2015oea}
\bibitem{Feng:2015oea}
X.~H.~Feng, H.~S.~Liu, H.~L\"u and C.~N.~Pope,
%``Black Hole Entropy and Viscosity Bound in Horndeski Gravity,''
JHEP \textbf{11} (2015), 176
doi:10.1007/JHEP11(2015)176
[arXiv:1509.07142 [hep-th]].
%60 citations counted in INSPIRE as of 30 Apr 2021


%\cite{Brito:2019ose}
\bibitem{Brito:2019ose}
F.~A.~Brito and F.~F.~Santos,
%``Black brane in asymptotically Lifshitz spacetime and viscosity/entropy ratios in Horndeski gravity,''
EPL \textbf{129} (2020) no.5, 50003
doi:10.1209/0295-5075/129/50003
[arXiv:1901.06770 [hep-th]].
%11 citations counted in INSPIRE as of 24 Dec 2021



%\cite{Kobayashi:2014eva}
\bibitem{Kobayashi:2014eva}
T.~Kobayashi and N.~Tanahashi,
%``Exact black hole solutions in shift symmetric scalar\textendash{}tensor theories,''
PTEP \textbf{2014} (2014), 073E02
doi:10.1093/ptep/ptu096
[arXiv:1403.4364 [gr-qc]].
%88 citations counted in INSPIRE as of 25 Jun 2021





\bibitem{Babichev:2013cya}
E.~Babichev and C.~Charmousis,
%``Dressing a black hole with a time-dependent Galileon,''
JHEP \textbf{08} (2014), 106
doi:10.1007/JHEP08(2014)106
[arXiv:1312.3204 [gr-qc]].
%242 citations counted in INSPIRE as of 23 Apr 2021




\bibitem{Anabalon:2013oea}
A.~Anabalon, A.~Cisterna and J.~Oliva,
%``Asymptotically locally AdS and flat black holes in Horndeski theory,''
Phys. Rev. D \textbf{89} (2014), 084050
doi:10.1103/PhysRevD.89.084050
[arXiv:1312.3597 [gr-qc]].
%164 citations counted in INSPIRE as of 23 Apr 2021


\bibitem{Rinaldi:2012vy}
M.~Rinaldi,
%``Black holes with non-minimal derivative coupling,''
Phys. Rev. D \textbf{86} (2012), 084048
doi:10.1103/PhysRevD.86.084048
[arXiv:1208.0103 [gr-qc]].
%179 citations counted in INSPIRE as of 23 Apr 2021


\bibitem{Bravo-Gaete:2014haa}
M.~Bravo-Gaete and M.~Hassaine,
%``Thermodynamics of a BTZ black hole solution with an Horndeski source,''
Phys. Rev. D \textbf{90} (2014) no.2, 024008
doi:10.1103/PhysRevD.90.024008
[arXiv:1405.4935 [hep-th]].
%32 citations counted in INSPIRE as of 23 Apr 2021



%\cite{Bravo-Gaete:2013dca}
\bibitem{Bravo-Gaete:2013dca}
M.~Bravo-Gaete and M.~Hassaine,
%``Lifshitz black holes with a time-dependent scalar field in a Horndeski theory,''
Phys. Rev. D \textbf{89} (2014), 104028
doi:10.1103/PhysRevD.89.104028
[arXiv:1312.7736 [hep-th]].
%54 citations counted in INSPIRE as of 28 Apr 2021



%\cite{BravoGaete:2019rci}
\bibitem{BravoGaete:2019rci}
M.~Bravo Gaete, S.~Gomez and M.~Hassaine,
%``Black holes with Lambert W function horizons,''
Eur. Phys. J. C \textbf{79} (2019) no.3, 200
doi:10.1140/epjc/s10052-019-6723-6
[arXiv:1901.09612 [hep-th]].
%5 citations counted in INSPIRE as of 30 Apr 2021







 %\cite{Smarr:1972kt}
\bibitem{Smarr:1972kt}
L.~Smarr,
%``Mass formula for Kerr black holes,''
Phys. Rev. Lett. \textbf{30} (1973), 71-73
[erratum: Phys. Rev. Lett. \textbf{30} (1973), 521-521]
doi:10.1103/PhysRevLett.30.71
%400 citations counted in INSPIRE as of 29 Jan 2021



  
\bibitem{Hajian:PRD2016}
K.~Hajian, M.~M.~sheikh-Jabbari, Phys.~Rev.~D {\bf 93}, 044074 (2016).

\bibitem{Hajian:PLB2020} 
K.~Hajian, S.~Liberati, M.~M.~Sheikh-Jabbari and M.~H.~Vahidinia, Phys.~Lett. B {\bf 812}, 136002 (2020).




\bibitem{Plebanski:1968}
  J.~Pleb\'anski,
  {\it Lectures on Non-Linear Electrodynamics} (Nordita, 1968).


\bibitem{Alvarez:2014pra}
A.~Alvarez, E.~Ay\'on-Beato, H.~A.~Gonz\'alez and M.~Hassa\"\i{}ne,
%``Nonlinearly charged Lifshitz black holes for any exponent $z>1$,''
JHEP \textbf{06} (2014), 041
doi:10.1007/JHEP06(2014)041
[arXiv:1403.5985 [gr-qc]].
%22 citations counted in INSPIRE as of 01 Feb 2021


\bibitem{Stetsko:2020nxb}
M.~M.~Stetsko,
%``Static spherically symmetric black hole in Einstein-power-Yang-Mills-dilaton theory and some aspects of its thermodynamics,''
[arXiv:2012.14902 [hep-th]].
%0 citations counted in INSPIRE as of 18 Feb 2021


\bibitem{Stetsko:2020tjg}
M.~M.~Stetsko,
%``Static dilatonic black hole with nonlinear Maxwell and Yang\textendash{}Mills fields of power-law type,''
Gen. Rel. Grav. \textbf{53} (2021) no.1, 2
doi:10.1007/s10714-020-02777-w
[arXiv:2012.14915 [hep-th]].
%1 citations counted in INSPIRE as of 18 Feb 2021






%\cite{BenAchour:2016fzp}
\bibitem{BenAchour:2016fzp}
J.~Ben Achour, M.~Crisostomi, K.~Koyama, D.~Langlois, K.~Noui and G.~Tasinato,
{\it Degenerate higher order scalar-tensor theories beyond Horndeski up to cubic order},
JHEP \textbf{12} (2016), 100,
[arXiv:1608.08135 [hep-th]].
%201 citations counted in INSPIRE as of 31 Dec 2020

%\cite{Motohashi:2016ftl}
\bibitem{Motohashi:2016ftl}
H.~Motohashi, K.~Noui, T.~Suyama, M.~Yamaguchi and D.~Langlois,
%``Healthy degenerate theories with higher derivatives,''
JCAP \textbf{07} (2016), 033
doi:10.1088/1475-7516/2016/07/033
[arXiv:1603.09355 [hep-th]].
%117 citations counted in INSPIRE as of 14 Jun 2021



\bibitem{Babichev:2020qpr}
E.~Babichev, C.~Charmousis, A.~Cisterna and M.~Hassaine,
%`` Regular black holes via the Kerr-Schild construction in DHOST theories,''
JCAP \textbf{06} (2020), 049
doi:10.1088/1475-7516/2020/06/049
[arXiv:2004.00597 [hep-th]].
%8 citations counted in INSPIRE as of 31 Dec 2020






%\cite{Baake:2020tgk}
\bibitem{Baake:2020tgk}
O.~Baake, M.~F.~Bravo Gaete and M.~Hassaine,
%`` Spinning black holes for generalized scalar tensor theories in three dimensions},''
Phys. Rev. D \textbf{102} (2020) no.2, 024088,
[arXiv:2005.10869 [hep-th]].
%1 citations counted in INSPIRE as of 31 Dec 2020


\bibitem{Baake:2021kyg}
O.~Baake and M.~Hassaine,
%``Rotating stealth black holes with a cohomogeneity-1 metric,''
Eur. Phys. J. C \textbf{81} (2021), 642
doi:10.1140/epjc/s10052-021-09449-2
[arXiv:2104.13834 [hep-th]].
%1 citations counted in INSPIRE as of 25 Dec 2021


%\cite{Baake:2021jzv}%\cite{Santos:2020lmb}
\bibitem{Baake:2021jzv}
O.~Baake, C.~Charmousis, M.~Hassaine and M.~San Juan,
%``Regular black holes and gravitational particle-like solutions in generic DHOST theories,''
JCAP \textbf{06} (2021), 021
doi:10.1088/1475-7516/2021/06/021
[arXiv:2104.08221 [hep-th]].
%10 citations counted in INSPIRE as of 25 Dec 2021

%\cite{Santos:2020lmb}
\bibitem{Santos:2020lmb}
F.~F.~Santos,
%``Complexity of four dimensional Anti-de-Sitter black holes with a rotating string in generalized scalar-tensor theories,''
[arXiv:2010.10942 [hep-th]].
%2 citations counted in INSPIRE as of 25 Dec 2021



\end{thebibliography}
\end{document}